\documentclass[aps,prd,showpacs,twocolumn,floatfix,nofootinbib,superscriptaddress,showkeys,longbibliography,
amsmath,amssymb,amsfonts]{revtex4-1} 

\usepackage{bm}
\usepackage{slashed}
\usepackage{graphicx}

\usepackage{amsmath,amssymb,amsfonts,graphicx}
\usepackage{dsfont, mathrsfs}
\usepackage{bm}
\usepackage{slashed}
\usepackage{subfigure}
\usepackage[svgnames]{xcolor}
\usepackage{placeins}
\usepackage{array}
\usepackage{natbib,hyperref}  
\usepackage{url}
\usepackage{lineno}
\usepackage[utf8]{inputenc}

\makeatletter
\newcommand{\thickhline}{%
    \noalign {\ifnum 0=`}\fi \hrule height 1pt
    \futurelet \reserved@a \@xhline
}
\newcolumntype{"}{@{\hskip\tabcolsep\vrule width 1pt\hskip\tabcolsep}}
\makeatother

\usepackage{xspace}

\newcolumntype{L}[1]{>{\raggedright\let\newline\\\arraybackslash\hspace{0pt}}m{#1}}
\newcolumntype{C}[1]{>{\centering\let\newline\\\arraybackslash\hspace{0pt}}m{#1}}
\newcolumntype{R}[1]{>{\raggedleft\let\newline\\\arraybackslash\hspace{0pt}}m{#1}}
\newcommand{\Eq}[1]{Eq.~(\ref{#1})}
\newcommand{\al}[1]{

\begin{align} #1 \end{align}}
\newcommand{\non}{\nonumber}
\newcommand{\vect}[1]{\boldsymbol{#1}}
\newcommand{\vh}[1]{\boldsymbol{\hat{#1}}}

\def\Tr{\mathrm{Tr}}

\def\SA{\sphericalangle}

\def\vf{\varphi}

\def\fR{\varphi_R}
\def\fRB{\varphi_{\bar{R}}}
\def\fL{\varphi_L}

\def\fK{\varphi_{k}}

\def \fk{\varphi_{k}}
\def \fkB{\varphi_{\bar{k}}}

\def\fKR{\varphi_{KR}}
\def\fKRB{\varphi_{\bar{K} \bar{R}}}
\def \fq{\varphi_{q}}

\def\kT{\vect{k}_T}
\def\kBT{\vect{\bar{k}}_T}
\def\RT{\vect{R}_T}

\def\RBT{\vect{\bar{R}}_T}
\def\qT{\vect{q}_T}
\def\b{\vect{b}}

\newcommand{\ImL}{1\columnwidth}
\newcommand{\GapCapt}{\vspace{-8pt}}

\begin{document}

\title{Semi-inclusive production of two back-to-back hadron pairs in $e^+e^-$ annihilation revisited} 

\preprint{ADP-17-30/T1036}

\author{Hrayr~H.~Matevosyan}
\thanks{ORCID: http://orcid.org/0000-0002-4074-7411}
\affiliation{ARC Centre of Excellence for Particle Physics at the Terascale,\\ 
and CSSM, Department of Physics, \\
The University of Adelaide, Adelaide SA 5005, Australia
\\ http://www.physics.adelaide.edu.au/cssm
}
\author{Alessandro~Bacchetta}
\thanks{ORCID: http://orcid.org/0000-0002-8824-8355}
\affiliation{Dipartimento di Fisica, Universit\`a degli Studi di Pavia,
 via Bassi 6, I-27100 Pavia, Italy, and}
\affiliation{INFN Sezione di Pavia, via Bassi 6, I-27100 Pavia, Italy}
\author{Dani\"el~Boer}
\thanks{ORCID: http://orcid.org/0000-0003-0985-4662}
\affiliation{Van Swinderen Institute for Particle Physics and Gravity, University of Groningen, Nijenborgh 4, NL-9747 AG Groningen, The Netherlands}
\author{Aurore~Courtoy}
\thanks{ORCID: http://orcid.org/0000-0001-8906-2440}
\affiliation{Instituto de F\'isica, Universidad Nacional Aut\'onoma de M\'exico, Apartado Postal 20-364, 01000 
Ciudad de M\'exico, M\'exico. }
\author{Aram~Kotzinian}
\thanks{ORCID: http://orcid.org/0000-0001-8326-3284}
\affiliation{Yerevan Physics Institute,
2 Alikhanyan Brothers St.,
375036 Yerevan, Armenia
}
\affiliation{INFN, Sezione di Torino, 10125 Torino, Italy
}
\author{Marco~Radici}
\thanks{ORCID: http://orcid.org/0000-0002-4542-9797}
\affiliation{INFN Sezione di Pavia, via Bassi 6, I-27100 Pavia, Italy}
\author{Anthony~W.~Thomas}
\thanks{ORCID: http://orcid.org/0000-0003-0026-499X}
\affiliation{ARC Centre of Excellence for Particle Physics at the Terascale,\\     
and CSSM, Department of Physics, \\
The University of Adelaide, Adelaide SA 5005, Australia
\\ http://www.physics.adelaide.edu.au/cssm
} 

\begin{abstract}
The cross section for back-to-back hadron pair production in $e^+e^-$ annihilation provides access to the dihadron fragmentation functions (DiFF) needed to extract nucleon parton distribution functions from the semi-inclusive deep inelastic scattering (SIDIS) experiments with two detected final state hadrons. Particular attention is given to the so-called interference DiFF (IFF), which makes it possible to extract the transversity parton distribution of the nucleon in the  collinear framework. However, previously unnoticed discrepancies were recently highlighted between the definitions of the IFFs appearing in the collinear kinematics when reconstructed from DiFFs entering the unintegrated fully differential cross sections of SIDIS and $e^+e^-$ annihilation processes. In this work, to clarify this problem we rederive the fully differential cross section for $e^+e^-$ annihilation at the leading-twist approximation. We find a mistake in the definition of the kinematics in the original expression that systematically affects a subset of terms and that leads to two significant consequences. First, the discrepancy between the IFF definitions in the cross sections for SIDIS and $e^+e^-$ annihilation is resolved. Second, the previously derived azimuthal asymmetry for accessing the helicity dependent DiFF $G_1^\perp$ in $e^+e^-$ annihilation vanishes, which explains the nonobservation of this asymmetry in the recent experimental searches by the {\tt BELLE} collaboration. We discuss the recently proposed alternative option to extract $G_1^\perp$.
\end{abstract}

\pacs{13.60.Hb,~13.60.Le,~13.87.Fh,~12.39.Ki}

\keywords{$e^+e^-$  to two hadron pairs, dihadron fragmentation functions}

\date{\today}                                           

\maketitle

\section{Introduction}
\label{SEC_INTRO}
 
The understanding of the complete spin-dependent structure of the nucleon has been at the forefront of studies in nuclear physics in recent decades. Particular attention has been given to studying the so-called transversity parton distribution function (PDF), which describes the correlation of the transverse polarization of the nucleon with the transverse polarization of its constituent partons (see e.g.~\cite{Barone:2001sp}). The chiral-odd nature of the transversity PDF makes it much harder to measure compared to the unpolarized and helicity dependent PDFs. Two approaches have been recently employed in phenomenological extractions of the transversity~\cite{Kang:2015msa,Anselmino:2015sxa,Bacchetta:2011ip,Pisano:2015wnq}. The first method uses the Collins effect~\cite{Collins:1992kk}, that describes the correlation between the transverse momentum of a produced hadron with the transverse polarization of an initial quark in the hadronization process. The convolution upon the transverse momenta of initial and final partons of the transversity and the Collins fragmentation function (FF) can be measured in a SIDIS process with a single measured final state hadron~\cite{Mulders:1995dh}, while the convolution of two Collins FFs are accessible from the semi-inclusive production of two back-to-back hadrons in $e^+e^-$ annihilation~\cite{Boer:1997mf}. The second method, based on DiFFs, leverages the correlation between the relative transverse momenta of two produced hadrons with the transverse polarization of a quark in its hadronization, which is quantified by the IFF $H^\SA_1$. Similarly to the previous method, here again the SIDIS process with two final state hadrons being measured is used to access a structure function containing the transversity PDF and an IFF~\cite{Bianconi:1999cd, Bianconi:1999uc, Radici:2001na,Bacchetta:2002ux}, while the semi-inclusive production of two back-to-back hadron pairs in $e^+e^-$ annihilation provides access to IFFs~\cite{Boer:2003ya,Bacchetta:2008wb,Courtoy:2012ry}. The advantage of the dihadron method compared to using the Collins effect is that it is possible to work in the collinear framework where the corresponding SIDIS structure function factorizes in a simple product of the transversity PDF and  the IFF, while for the single hadron case the transversity is convoluted with the Collins function via an integral involving their transverse momentum dependences. The same is true for the structure functions containing the IFF and the Collins FF, respectively, in the $e^+e^-$ annihilation cross section. Moreover, in the collinear framework the same combination of transversity PDF and IFF can be explored also in proton-proton collisions leading to the semi-inclusive production of dihadron pairs~\cite{Bacchetta:2004it,Radici:2016lam}, while this possibility is in principle precluded for the Collins effect due to factorization breaking contributions. Finally, the evolution equations  connecting the IFF at different scales of the various processes have a simple standard form~\cite{Ceccopieri:2007ip}, while the evolution of a transverse-momentum dependent PDF is more complicated and depends on non perturbative parameters~\cite{Collins:2011zzd}.

A major experimental effort to measure the various azimuthal asymmetries involved in extracting the transversity PDF using the dihadron way has been made by several collaborations, such as {\tt HERMES}~\cite{Airapetian:2008sk}, {\tt COMPASS}~\cite{Adolph:2012nw,Adolph:2014fjw}, and {\tt BELLE}~\cite{Vossen:2011fk,Seidl:2017qhp}. The IFFs from $e^+e^-$ measurements at {\tt BELLE} were fitted in Refs.~\cite{Courtoy:2012ry,Radici:2015mwa}. In turn these were used in Refs.~\cite{Bacchetta:2011ip,Bacchetta:2012ty,Radici:2015mwa} to successfully extract the transversity PDF using {\tt HERMES} and {\tt COMPASS} data. Recently, the {\tt STAR} collaboration released also dihadron data for azimuthal asymmetries in proton-proton collisions with a transversely polarized proton~\cite{Adamczyk:2015hri,Adamczyk:2017ynk} which can be included in an attempt of extracting the transversity PDF from a global fit~\cite{Radici:2017qez}.
 
 Recently, systematic model calculations of both FFs and DiFFs for unpolarized hadrons have been performed within the extended quark-jet model, which for the first time provides a self-consistent description for the hadronization of a quark with an arbitrary polarization~\cite{Bentz:2016rav,Matevosyan:2016fwi, Matevosyan:2017alv, Matevosyan:2017uls}. The two DiFFs, $H^\SA_1$ and $H^\perp_1$, describing the correlations between the relative and the total transverse moment of the hadron pair with the transverse polarization of the quark, respectively, were studied in Ref.~\cite{Matevosyan:2017uls}. 
There, it was observed that the integrated IFF built from the DiFFs entering the unintegrated SIDIS cross section is different from the one that is built from the corresponding unintegrated cross section for $e^+e^-$ annihilation derived in~\cite{Boer:2003ya}. In particular, in SIDIS the integrated IFF contains both the zeroth Fourier cosine moment of the fully unintegrated $H^\SA_1$, along with the first Fourier cosine moment of $H^\perp_1$. This admixture of $H^\perp_1$ did not appear in the original derivation in Ref.~\cite{Radici:2001na} but was later included in Ref.~\cite{Bacchetta:2003vn}. On the other hand, the integrated $H^\SA_1$ in $e^+e^-$ annihilation in Ref.~\cite{Boer:2003ya} contains only the zeroth Fourier cosine moment of the unintegrated $H^\SA_1$. The model estimates of these two definitions of IFFs in Ref.~\cite{Matevosyan:2017uls} produced almost a factor of two discrepancy between them. 

Another prediction of Ref.~\cite{Boer:2003ya} concerned a particular azimuthal modulation that provides access to the first Fourier cosine moment of the quark helicity dependent DiFF $G_1^\perp$. However, the recent preliminary results from the {\tt BELLE}  collaboration showed no signal for this modulation within the experimental  uncertainties~\cite{Abdesselam:2015nxn, Vossen:2015znm}.  The recent {\tt COMPASS} studies~\cite{Sirtl:2017rhi} also yielded no significant signal for SIDIS. Even though the model calculations of Ref.~\cite{Matevosyan:2017alv} suggest that the integrated $G_1^\perp$ appearing in Ref.~\cite{Boer:2003ya}  is naturally smaller in magnitude than the $H^\SA_1$, this was still a surprise given the precision achieved in the {\tt BELLE} analysis.

In this work, we rederive the unintegrated cross section for the semi-inclusive production of two back-to-back hadron pairs in $e^+e^-$ annihilation, first performed in Ref.~\cite{Boer:2003ya}. We then recalculate the azimuthal asymmetries used for extracting the IFFs and the helicity dependent DiFF in order to resolve the above discrepancies.

 This paper is organized in the following way.  In the next section we briefly review the formalism for DiFFs. In Sec.~\ref{SEC_EE_XSEC}, we describe the kinematics of two hadron pair production in $e^+e^-$ annihilation and rederive the corresponding cross section. In Sec.~\ref{SEC_EE_ASYMM},  we rederive both azimuthal asymmetries involving  $H^\SA_1$ and  $G^\perp_1$. We present our conclusions in Sec.~\ref{SEC_CONCLUSIONS}.

\section{Field-theoretical definitions of the DiFFs}
\label{SEC_DIFF_FORM}
 
   The fragmentation of a quark $q$ of an arbitrary polarization $\vect{s}$ into two unpolarized hadrons $h_1, h_2$ is fully described at the leading twist approximation by four DiFFs, see Refs.~\cite{Bianconi:1999cd, Bianconi:1999uc, Radici:2001na,Bacchetta:2003vn, Boer:2003ya}. The relevant kinematics is described by the momentum $k$ and mass $m$ of the quark $q$, and the  corresponding momenta $P_1, P_2$ and masses $M_1,M_2$ of the $h_1,h_2$ pair. In the definitions of the DiFFs, the momenta $P_1$ and $P_2$ of the individual hadrons are replaced by their total, $P \equiv P_h$, and relative, $R$, momenta
\al
{
\label{EQ_P_TOT}
& P \equiv P_h = P_1 + P_2,
\\
\label{EQ_P_REL}
& R = \frac{1}{2}( P_1 - P_2),
}
with $P_h^2 = M_h^2$ the squared invariant mass of the pair.
 
\begin{figure}[t]
\centering 
\includegraphics[width=\ImL]{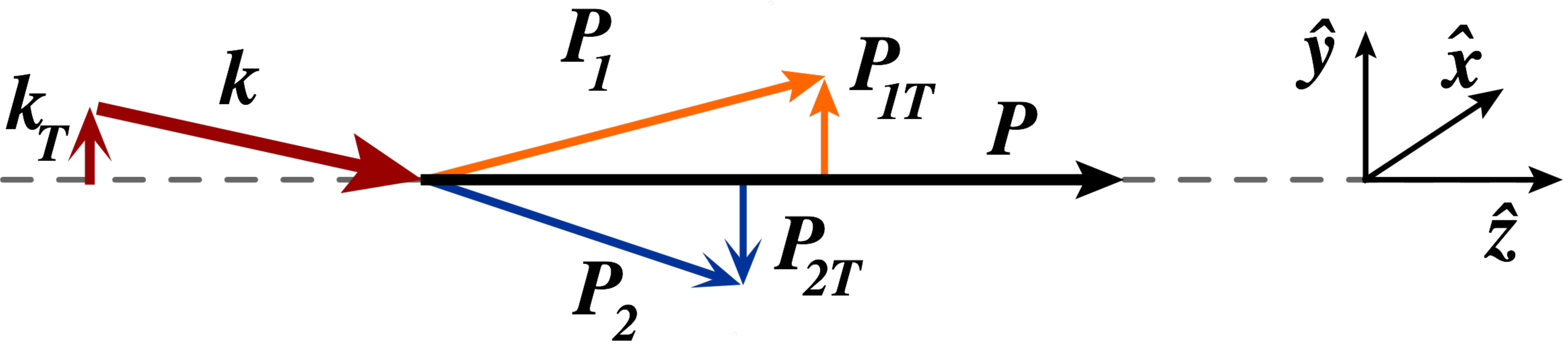}
\GapCapt
\caption{The dihadron fragmentation coordinate system, where  the  $\hat{z}$ axis is taken along the total 3-momentum of the two hadrons, $\vect{P}$. The components of 3-momenta perpendicular to $\hat{z}$ axis are denoted with a subscript $_T$.}
\label{PLOT_DIFF_KT_SYS}
\end{figure}

The $\hat{z}$ axis is defined along the spatial component of the total momentum $P_h$ and the components of three-vectors perpendicular to the $\hat{z}$ direction are denoted by subscript $_T$, as schematically shown in Fig.~\ref{PLOT_DIFF_KT_SYS}. 

The light-cone momentum fractions of the hadrons are defined as the ratios of the plus components\footnote{The light-cone components of a 4-vector $a$ are defined as $a = (a^+,a^-,\vect{a}_T)$, where $a^\pm = \frac{1}{\sqrt{2}}(a^0 \pm a^3)$} of their four vectors to the quark momentum, $z_i = {P_i^+}/{k^+}$.  The following light-cone momentum fractions are used in the definitions of the DiFFs
\al
{
 z &= z_1  + z_2,
\\
 \xi &= \frac{z_1}{z}  = 1- \frac{z_2}{z} \, .
}

The two-hadron fragmentation of a quark is described by a quark-quark correlator~\cite{Bianconi:1999cd, Radici:2001na, Boer:2003ya, Boer:2008fr}
\al
{
\Delta_{ij}(k; P_h, R)& 
\\ \non
 = \sum_X \int d^4 \zeta &e^{i k\cdot \zeta} \langle 0 |\psi_i(\zeta) | P_h R, X \rangle \langle P_h R, X | \bar{\psi}_j(0) | 0 \rangle,
 }
which, for the case of unpolarized hadron pair and at the leading twist approximation, is parametrized  via four DiFFs
\al
{
\frac{1}{32z} \int d k^- \Delta(k, &P_h, R) |_{k^+ = P_h^+/z} \equiv \Delta(z, \xi,\kT, \RT) 
\\ \non
 =\frac{1}{4\pi}\, \frac{1}{4}\Bigg\{
 & D_1 \slashed{n}_{+}
  - G_1^\perp\frac{\epsilon_{\mu\nu\rho\sigma} \gamma^\mu n_+^\nu k_T^\rho R_T^\sigma}{M_h^2} \gamma_5
\\ \non
&
 + H_1^\SA\frac{\sigma_{\mu\nu} R_T^\mu n_+^\nu}{M_h}
 + H_1^\perp\frac{\sigma_{\mu\nu} k_T^\mu n_+^\nu}{M_h}
 \Bigg\},
}
where $D_1$ is the unpolarized DiFF, $G_1^\perp$ is the helicity dependent DiFF, $H_1^\SA$ is the IFF, and $H_1^\perp$ is the analogue of the Collins function for the dihadron case.  The lightlike vectors $n_-$ and $n_+$ are defined as for any 4-vector $a$, namely $a^\pm = a \cdot n_\mp$, and $n_+ n_- = 1$, $ n_+^2 = n_-^2 =0$. All four DiFFs are functions of $z, \xi, |\kT|, |\RT|$, and $\kT~\cdot~ \RT = |\kT| |\RT| \cos(\fk - \fR)$, where $\fR$  and $\fK$ denote the azimuthal angles of the vectors $\RT$ and $\kT$. Thus, the DiFFs only depend on the cosine of the difference of the azimuthal  angles $\fk-\fR$, that we denote as $\fKR$.  The DiFFs can be further expanded in an infinite series of Fourier moments with respect to angle $\fKR$, as done in Ref.~\cite{Matevosyan:2017uls} (see also Ref.~\cite{Gliske:2014wba} for an alternative expansion). It is clear, that all the sine terms vanish, as the DiFFs are even functions of $\fKR$. 

For $D_1$ we have
\al
{
 \label{EQ_FOURIER_FRK}
D_1 (z, \xi, \kT^2, \RT^2,& \cos(\fKR)) 
\\ \non
  =& \frac{1}{\pi} \sum_{n=0}^\infty
\frac{\cos(n \cdot \fKR)}{1+\delta_{0,n}} \ D_1^{[n]}(z, \xi, |\kT|, |\RT| ),
}
and similarly for the other DiFFs.

The invariant mass of the hadron pair $M_h$ is used to replace the magnitude of $\RT$
\al
{
 R_T^2 & =  \xi (1-\xi) M_h^2 - M_1^2 (1-\xi) - M_2^2 \xi.
}

These Fourier decompositions will prove valuable when examining the azimuthal dependence of various structure functions of the $e^+e^-$ cross section which we rederive in the next section. 

%
\section{The $e^+e^-$  cross section}
\label{SEC_EE_XSEC}

In this section we rederive the $e^+e^-\to h_1 h_2 + \bar{h}_1 \bar{h}_2 + X$ cross section at the leading twist approximation, following the framework set out in the original work of Boer {\em et al.}~\cite{Boer:1997mf,Boer:1997qn,Boer:2003ya}. First, we briefly lay out the kinematics in the next subsection, followed by the evaluation of the cross section itself in the subsequent subsection.

\subsection{Kinematics}
\label{SUBSEC_EE_KINEMATICS}

\begin{figure}[t]
\centering 
\vspace{-1cm}
\includegraphics[width=\ImL]{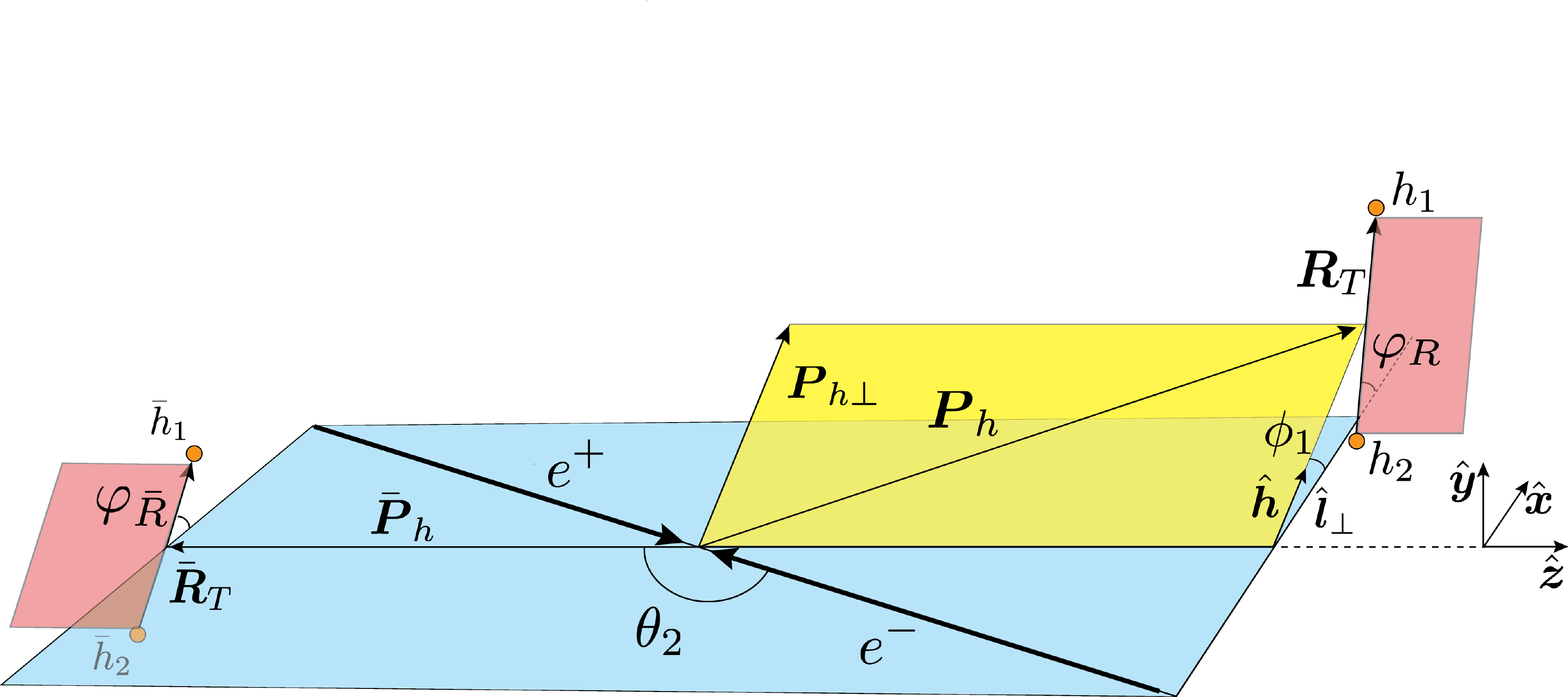}
\GapCapt
\caption{The kinematics of $e^+e^-$ annihilation.}
\label{PLOT_EE_KINEMATICS}
\end{figure}

A schematic depiction of the kinematic setup is shown in Fig.~\ref{PLOT_EE_KINEMATICS}. Here, the electron with momentum $l$ annihilates with a positron of momentum $l'$, creating a quark-antiquark pair. The time like momentum of the intermediate boson in this hard process is denoted as $q=l+l'$ and we define $q^2=Q^2$. In this work we use $Q$ as the hard scale and will ignore all the contributions of order $1/Q$. The quark and antiquark hadronize, producing two back-to-back jets. We choose a hadron pair $h_1,h_2$ with momenta $P_1,P_2$ and masses $M_1,M_2$ from one of the jets. From the other jet, we choose the second hadron pair $\bar{h}_1,\bar{h}_2$, with momenta  $\bar{P}_1,\bar{P}_2$ and masses $\bar{M}_1,\bar{M}_2$. Here again we define the total and relative transverse momenta for each pair, as done in Eqs.~(\ref{EQ_P_TOT},\ref{EQ_P_REL}), and denote the corresponding momenta for the $\bar{h}_1,\bar{h_2}$ pair as $\bar{P}_h$ and $\bar{R}$. In the "leading hadron approximation", where we assume that a significant fraction of the energy in each jet is carried by the two pairs, we can  write $P_h \cdot \bar{P}_h \sim Q^2$. Then we decompose the momenta $P_h$, $\bar{P}_h$ and $q$ in light-cone coordinates in a frame where $\vect{P}_{hT}=0$ and $\vect{\bar{P}}_{hT}=0$, to identify the corresponding dominant terms
\al
{
\label{EQ_PH_LC}
 P_h^\mu &= \frac{M_h^2}{z_h Q \sqrt{2}} n_-^\mu + \frac{z_h Q}{\sqrt{2}} n_{+}^\mu  \approx \frac{z_h Q}{\sqrt{2}} n_+^\mu,
 \\
 \label{EQ_PHB_LC}
 \bar{P}_h^\mu &= \frac{\bar{z}_h Q}{\sqrt{2}} n_{-}^\mu + \frac{\bar{M}_h^2}{\bar{z}_h Q \sqrt{2}} n_+^\mu \approx \frac{\bar{z}_h Q}{\sqrt{2}} n_-^\mu,
 \\
 \label{EQ_Q_LC}
 q^\mu &= \frac{ Q}{\sqrt{2}} n_{-}^\mu + \frac{Q}{\sqrt{2}} n_+^\mu  + q_T^\mu,
 }
where
\al
{
 z_h &= \frac{2 P_h\cdot q}{Q^2} \approx z,
 \\
 \bar{z}_h &= \frac{2 \bar{P}_h\cdot q}{Q^2} \approx \bar{z},
}
and 
\al
{
  - q_T^2 = Q_T^2 \ll Q^2.
}

 We can project the components of 4-vectors transverse to $n_\pm$  using the tensors
\al
{
g_T^{\mu \nu} &= g^{\mu\nu} - n_+^\mu n_-^\nu - n_+^\nu n_-^\mu,
\\
\epsilon_T^{\mu \nu} &=  \epsilon^{\mu \nu \rho \sigma} n_{+ \rho} n_{- \sigma},
}
where $g^{\mu\nu}$ is the metric tensor.

 The coordinate system in Fig.~\ref{PLOT_EE_KINEMATICS} is defined by taking the $\hat{z}$ axis opposite to the 3-momentum $\vect{\bar{P}}_h$, while the components of the vectors perpendicular to $\hat{z}$ are denoted with a subscript $_\perp$ in a frame where $\vect{q}_\perp=0$. It can be easily shown, that $\vect{P}_{h\perp} = - {z} \vect{q}_T$, up to negligible correction of order $Q_T^2/Q^2 \ll 1$. We can then define the two orthogonal unit vectors in $\perp$ direction 
\al
{
 \vh{h} &= \frac{\vect{P}_{h\perp}}{|\vect{P}_{h\perp}|}  = -\frac{ \vect{q}_T}{| \vect{q}_T|},
\\
\hat{g}^i &= \epsilon_T^{ij} \hat{h}^j =  \epsilon^{0ij3} \hat{h}^j,
}
where the following convention is used $\epsilon^{0123} = +1$. 

To keep consistency, we will define all the azimuthal angles with respect to the lepton frame. Then, we can parametrize these two vectors using the azimuthal angle $\phi_1$ of $\vh{h}$
\al
{
\vh{h} & = (\cos(\phi_1), \sin(\phi_1)),
\\
\vh{g} &= (\sin(\phi_1), -\cos(\phi_1)),
}
so that the azimuthal angle of $\vh{g}$ is simply $\phi_g = 3/2\pi + \phi_1$.

The lepton plane in Fig.~\ref{PLOT_EE_KINEMATICS} is spanned by the $\hat{z}$ axis and the transverse component $\vect{l}_\perp$ of the electron momentum $l$. The unit vector $\vh{l}_\perp$ can be parametrized using the lepton plane angle $\fL$ in the laboratory frame. However, all the following results are independent of the orientation of the scattering plane with respect to the laboratory frame, hence the $\fL$ dependence will be ignored. Here we can also define the associated normalized 4-vector
\al
{
 \hat{l}_\perp = \frac{l_\perp}{|\vect{l}_\perp|}.
}

Similar to the light-cone frame, we can now define a set of orthogonal normalized 4-vectors 
\al
{
 \hat{t} &= \frac{q}{Q},
 \\
 \hat{v} &= 2 \frac{\bar{P}_{h}}{\bar{z} Q} - \hat{t},
}
where the spacelike vector $\hat{v}$ is denoted as $\hat{z}$ in Refs.~\cite{Boer:1997mf,Boer:1997qn,Boer:2003ya}. Here we changed the notation to avoid any possible confusion with the notation of the $\vh{z}$ axis. The orthogonal projections of the 4-vectors can be again achieved using the tensors
\al
{
g_\perp^{\mu\nu} &= g^{\mu\nu} - \hat{t}^\mu \hat{t}^\nu + \hat{v}^\nu \hat{v}^\mu,
\\
\epsilon_\perp^{\mu \nu} &= - \epsilon^{\mu \nu \rho \sigma}  \hat{t}_\rho \hat{v}_\sigma,
}

The two perpendicular projection tensors can be related
\al
{
g_\perp^{\mu\nu} &= g_T^{\mu\nu} - \frac{n_+^\mu q_T^\nu + n_+^\nu q_T^\mu}{Q}.
}
In this work we neglect all terms of order $Q_T/Q$, $M_h/Q$, $\bar{M}_h/Q$. Thus we also neglect the differences between the $_T$ and $_\perp$ components of vectors.

\subsection{Cross section}
\label{SUBSEC_EE_XSEC}

The cross section for this process is given by the convolution of leptonic and hadronic tensors
\al
{
 \frac{2 P_1^0 2P_2^0 2\bar{P}_1^0 2\bar{P}_2^0 d \sigma}
 {d^3 \vect{P}_1  d^3 \vect{P}_2  d^3 \vect{\bar{P}}_1  d^3 \vect{\bar{P}}_2 }
 = \frac{\alpha^2}{Q^6} L_{\mu\nu} W^{\mu\nu}_{(4h)},
}
where
\al
{
\label{EQ_LEPT_TENS}
 L_{\mu\nu} = Q^2\Bigg[ 
&-2 A(y) g_\perp^{\mu\nu} 
\\ \non 
&+ 4B(y) \hat{v}^\mu \hat{v}^\nu - 4B(y)\Big( \hat{l}_\perp^\mu \hat{l}_\perp^\nu + \frac{1}{2} g_\perp^{\mu\nu} \Big)
\\ \non 
&- 2C(y) B^{1/2}(y) \Big( \hat{v}^\mu \hat{l}_\perp^\nu + \hat{v}^\nu \hat{l}_\perp^\mu \Big) 
 \Bigg],
}
and
\al
{
 A(y) &= \frac{1}{2} - y + y^2,
\\
B(y) &= y(1-y),
\\
C(y) &= 1-2y,
}
with 
\al
{
\label{EQ_Y_TH2}
 y = \frac{P_h \cdot l}{P_h \cdot q} \approx \frac{l^-}{q^-} = \frac{1+ \cos \theta_2}{2}.
}
The last equality holds in the center-of-mass frame, where $\theta_2$ is the angle between the 3-momentum of the electron $l$ and the $\vh{z}$ axis.

\begin{widetext}

The hadronic tensor is defined as 
\al
{
W^{\mu\nu}_{(4h)}(q; P_h, R, \bar{P}_h, \bar{R}) 
 = \frac{1}{(2\pi)^{10}}\sum_X  \int  \frac{d^3 \vect{P}_X}{(2\pi)^3 2 P_X^0} 
 \ (2\pi)^4 \ & \delta(q - P_X - P_h - \bar{P}_h )
\\ \non
& \times \langle 0| J^\mu(0) | P_X; P_h, R, \bar{P}_h , \bar{R} \rangle
 \langle  P_X; P_h, R, \bar{P}_h , \bar{R} | J^\nu(0) | 0 \rangle .
}

Using the parton picture, we can decompose the hadronic tensor in terms of the quark-quark correlators $\Delta$ and $\bar{\Delta}$ for the production of the two hadron pairs in the fragmentation of the quark and the antiquark
\al
{
\label{EQ_W_TENS_D_DBAR}
 W^{\mu\nu}_{(4h)} \approx & \ 3 (32 z) (32\bar{z})
 \sum_{a, \bar{a}} e_a^2
 \int d^2 \kT d^2 \kBT \ \delta^2(\qT - \kT - \kBT)
 \Tr \Bigg[ \bar{\Delta}(\bar{z},\bar{\xi}, \kBT, \RBT) \gamma^\mu 
\Delta(z,\xi, \kT, \RT) \gamma^\mu
\Bigg],
}
where $a$ denotes the flavor of the fragmenting quark and the prefactor is the number of active colors $N_C=3$. 

Following the transformation of the phase space factor detailed in Ref.~\cite{Boer:2003ya}, the cross section expression can be written as
\al
{
 \frac{d \sigma}{d^2 \qT\ dz\ d\xi\ d\fR\ d M_h^2\ d \bar{z}\ d\bar{\xi}\ d\fRB\ d \bar{M}_h^2\ dy \ d \fL} &
=\frac{\alpha^2}{128 Q^4} z \bar{z} L_{\mu\nu} W^{\mu\nu}.
}

Up until this point we have followed the same formalism and definitions as in Ref.~\cite{Boer:2003ya}. The next step is to evaluate the trace in \Eq{EQ_W_TENS_D_DBAR} and contract the resulting expression for the hadronic tensor with the leptonic tensor in \Eq{EQ_LEPT_TENS}. The resulting expression follows

\al
{
\label{EQ_XSEC_V1}
&\frac{d \sigma \Big( e^+e^- \to (h_1 h_2) (\bar{h}_1 \bar{h}_2) X \Big)}{d^2 \qT dz d\xi d\fR d M_h^2 d \bar{z} d\bar{\xi} d\fRB d \bar{M}_h^2 dy } 
 =  \frac{3 \alpha^2}{\pi Q^2} z^2 \bar{z}^2 \sum_{a,\bar{a}} e_a^2 
\\ \non
& \hspace{1cm}
\times
\Bigg\{
\hspace{0.2cm}
A(y) \mathcal{F}\Bigg[  D_1^a \bar{D}_1^{\bar{a}}\Bigg]
+ \cos(\fR + \fRB ) B(y) \frac{|\RT|}{M_h} \frac{|\RBT|}{\bar{M}_h} \mathcal{F}\Bigg[H_1^{\SA a} \bar{H}_1^{\SA \bar{a}}\Bigg]
\\ \non
& \hspace{2cm} + \cos(2\phi_1) \frac{B(y)}{M_h \bar{M}_h} \mathcal{F}\Bigg[\Big(2 (\vh{h}\cdot \kT) (\vh{h}\cdot \kBT) - (\kT \cdot \kBT) \Big) H_1^{\perp a} \bar{H}_1^{\perp \bar{a}}\Bigg]
\\ \non
&  \hspace{2cm} + \sin(2\phi_1 )\frac{B(y)}{M_h \bar{M}_h} \mathcal{F}\Bigg[\Big((\vh{g}\cdot \kT) (\vh{h}\cdot \kBT) + (\vh{h}\cdot \kT) (\vh{g}\cdot \kBT) \Big) H_1^{\perp a} \bar{H}_1^{\perp \bar{a}}\Bigg]
\\ \non
&  \hspace{2cm} + \cos(\phi_1 + \fR ) \frac{B(y) |\RT|}{M_h \bar{M}_h} \mathcal{F}\Bigg[ (\vh{h} \cdot \kBT)   H_1^{\SA a} \bar{H}_1^{\perp \bar{a}}\Bigg]
 + \sin(\phi_1 + \fR ) \frac{B(y) |\RT|}{M_h \bar{M}_h} \mathcal{F}\Bigg[ (\vh{g} \cdot \kBT)   H_1^{\SA a} \bar{H}_1^{\perp \bar{a}}\Bigg]
\\ \non
&  \hspace{2cm} +  \cos(\phi_1 + \fRB) \frac{B(y)  |\RBT|}{M_h \bar{M}_h} \mathcal{F}\Bigg[(\vh{h} \cdot \kT) H_1^{\perp a} \bar{H}_1^{\SA \bar{a}}\Bigg]
+  \sin(\phi_1 + \fRB ) \frac{B(y) |\RBT|}{M_h \bar{M}_h} \mathcal{F}\Bigg[(\vh{g} \cdot \kT) H_1^{\perp a} \bar{H}_1^{\SA \bar{a}}\Bigg]
\\ \non 
&\hspace{1.6cm} - A(y)  \frac{|\RT|}{M_h^2} \frac{|\RBT|}{\bar{M}_h^2} \Bigg(
 \hspace{0.4cm}   \sin(\phi_1 - \fR)   \sin(\phi_1 - \fRB) \mathcal{F}\Bigg[ \Big((\vh{h}\cdot \kT) (\vh{h}\cdot \kBT) \Big) G_1^{\perp a} \bar{G}_1^{\perp \bar{a}} \Bigg]
\\ \non
&\hspace{4.7cm} - \sin(\phi_1 - \fR) \cos(\phi_1 - \fRB) \mathcal{F}\Bigg[ \Big((\vh{h}\cdot \kT) (\vh{g}\cdot \kBT) \Big) G_1^{\perp a} \bar{G}_1^{\perp \bar{a}} \Bigg]
\\ \non
&\hspace{4.7cm} - \cos(\phi_1 - \fR) \sin(\phi_1 - \fRB) \mathcal{F}\Bigg[ \Big((\vh{g}\cdot \kT) (\vh{h}\cdot \kBT) \Big) G_1^{\perp a} \bar{G}_1^{\perp \bar{a}} \Bigg]
\\ \non
& \hspace{4.7cm} + \cos(\phi_1 - \fR) \cos(\phi_1 - \fRB) \mathcal{F}\Bigg[ \Big((\vh{g}\cdot \kT) (\vh{g}\cdot \kBT) \Big) G_1^{\perp a} \bar{G}_1^{\perp \bar{a}} \Bigg]
\Bigg)
\Bigg\} ,
}
where the convolution $\mathcal{F}$ is defined as
\al
{
\label{EQ_F_CONVOL}
\mathcal{F}[w D^a \bar{D}^{\bar{a}} ] =  \int d^2 \kT d^2 \kBT
   \delta^2(\vect{k}_T + \vect{\bar{k}}_T - \vect{q}_T)w( \kT, \kBT, \RT, \RBT)
 D^a(z, \xi, \kT^2, \RT^2, \kT \cdot \RT  )
 D^{\bar{a}}(\bar{z}, \bar{\xi}, \kBT^2, \RBT^2, \kBT \cdot \RBT).
}

There are several important differences between the expression in \Eq{EQ_XSEC_V1} and the original expression in Eq.~(19) of Ref.~\cite{Boer:2003ya}, apart from the different mass normalization. First, the terms multiplying $\vh{g}$ are multiplied by a factor of $-1$ in our expression. Second, the factor $A(y)$ in front of the $G_1^{\perp a} \bar{G}_1^{\perp \bar{a}} $ terms is also multiplied by a factor of $-1$. Lastly, the dependence on angle $\fL$ vanishes altogether, as in this work all the azimuthal angles are defined with respect to the lepton plane.

These differences allow us to rewrite the cross section in a much more compact form
\al
{
\label{EQ_XSEC_V2}
&\frac{d \sigma \Big( e^+e^- \to (h_1 h_2) (\bar{h}_1 \bar{h}_2) X \Big)}{d^2 \qT dz d\xi d\fR d M_h^2 d \bar{z} d\bar{\xi} d\fRB 
d \bar{M}_h^2 dy } 
 =  \frac{3 \alpha^2}{\pi Q^2} z^2 \bar{z}^2 \sum_{a,\bar{a}} e_a^2 
\Bigg\{
 A(y)  \mathcal{F}\Bigg[  D_1^a \bar{D}_1^{\bar{a}}\Bigg]
\\ \non
&\hspace{3cm}  + B(y) \mathcal{F}\Bigg[ \frac{|\kT|}{M_h} \frac{|\kBT|}{\bar{M}_h} \cos(\fK +\fkB ) H_1^{\perp a} \bar{H}_1^{\perp \bar{a}}\Bigg]
+ B(y) \mathcal{F}\Bigg[ \frac{|\RT|}{M_h} \frac{|\RBT|}{\bar{M}_h} \cos(\fR +\fRB) H_1^{\SA a} \bar{H}_1^{\SA \bar{a}}\Bigg]
\\ \non
&\hspace{3cm} + B(y) \mathcal{F}\Bigg[ \frac{|\kT|}{M_h} \frac{|\RBT|}{\bar{M}_h} \cos(\fk +\fRB) H_1^{\perp a} \bar{H}_1^{\SA \bar{a}}\Bigg]
+ B(y) \mathcal{F}\Bigg[ \frac{|\RT|}{M_h} \frac{|\kBT|}{\bar{M}_h} \cos(\fR +\fkB) H_1^{\SA a} \bar{H}_1^{\perp \bar{a}}\Bigg]
\\ \non 
&\hspace{3cm} - A(y) \mathcal{F}\Bigg[ \frac{|\RT| \, |\kT|}{M_h^2} \frac{|\RBT| \, |\kBT|}{\bar{M}_h^2} \sin(\fk -\fR) \sin(\fkB -\fRB)
  G_1^{\perp a} \bar{G}_1^{\perp \bar{a}} \Bigg]
\ \Bigg\}.
}

We obtain the cross section in collinear kinematics by integrating upon $d^2 \qT$. This integration trivially breaks up the convolution between $\kT$ and $\kBT$ in \Eq{EQ_F_CONVOL}. In the last line, we have the product of two terms of the following form
\al
{
\int d \fk \sin(\fKR) G_1^{\perp a}(z,\xi, |\kT|, |\RT|, \cos(\fKR)) = 0,
}
that trivially vanishes by changing the integration variable $\fk \to \fKR$. By replacing in \Eq{EQ_XSEC_V2} the remaining DiFFs with their Fourier cosine decompositions in  \Eq{EQ_FOURIER_FRK}, we have
\al
{
\label{EQ_XSEC_COLL}
&\frac{d \sigma \Big( e^+e^- \to (h_1 h_2) (\bar{h}_1 \bar{h}_2) X \Big)}{dz d\xi d\fR d M_h^2 d \bar{z} d\bar{\xi} d\fRB d\bar{M}_h^2 dy } 
 =  \frac{3 \alpha^2}{\pi Q^2} z^2 \bar{z}^2 \int d^2 \kT  \int d^2 \kBT \, \sum_{n,m}  
 \frac{\cos(n \fKR)}{\pi (1+\delta_{0,n})}  \frac{\cos(m \fKRB)}{\pi (1+\delta_{0,m})} \sum_{a,\bar{a}} e_a^2
 \\ \non
& \hspace{0.1cm}
 \times \Bigg\{
A(y) D_1^{a, [n]}  \bar{D}_1^{\bar{a}, [m]}  
+\frac{B(y)}{M_h \bar{M}_h } \Bigg(
 \cos(\fK +\fkB)  |\kT|  H_1^{\perp a, [n]} |\kBT| \bar{H}_1^{\perp \bar{a}, [m]} 
+  \cos(\fR +\fRB ) |\RT| H_1^{\SA a, [n]} |\RBT| \bar{H}_1^{\SA \bar{a}, [m]}
\\ \non
&\hspace{4.8cm} 
+ \cos(\fk + \fRB ) |\kT| H_1^{\perp a, [n]} |\RBT| \bar{H}_1^{\SA \bar{a}, [m]}  
+  \cos(\fkB  + \fR) |\RT| H_1^{\SA a, [n]}  |\kBT| \bar{H}_1^{\perp \bar{a}, [m]} 
\Bigg)
\Bigg\}
\\ \non
& =  \frac{3 \alpha^2}{\pi Q^2} z^2 \bar{z}^2 \int \frac{d \kT^2}{2}  \int \frac{d \kBT^2}{2} \, \sum_{a,\bar{a}} e_a^2
\Bigg\{
A(y) D_1^{a, [0]}  \bar{D}_1^{\bar{a}, [0]}  
+\frac{B(y)}{M_h \bar{M}_h }  \cos(\fR +\fRB) \Big(
  |\kT|  H_1^{\perp a, [1]} |\kBT| \bar{H}_1^{\perp \bar{a}, [1]} 
\\ \non
&\hspace{4.7cm} 
+  |\RT| H_1^{\SA a, [0]} |\RBT| \bar{H}_1^{\SA \bar{a}, [0]}
+  |\kT| H_1^{\perp a, [1]} |\RBT| \bar{H}_1^{\SA \bar{a}, [0]}  
+  |\RT| H_1^{\SA a, [0]}  |\kBT| \bar{H}_1^{\perp \bar{a}, [1]} 
\Big)
 \Bigg\}
\\ \non
&\hspace{0.2cm} 
= \frac{3 \alpha^2}{4\pi^3 Q^2}  \sum_{a,\bar{a}} e_a^2  \Bigg\{ A(y) D_1^a (z, \xi, M_h^2) \bar{D}_1^{\bar{a}} (\bar{z}, \bar{\xi}, \bar{M}_h^2) 
+ B(y)  \cos(\fR +\fRB ) \frac{|\RT|}{M_h} \frac{|\RBT|}{\bar{M}_h} H_1^{\SA a} (z, \xi, M_h^2) \bar{H}_1^{\SA \bar{a}} (\bar{z}, \bar{\xi}, \bar{M}_h^2) 
\Bigg\} , 
}
where
\al
{
\label{EQ_D_COLL}
D_1 (z, \xi, M_h^2) &\equiv z^2 \int d^2\kT \, D_1^{[0]} (z, \xi, |\kT|, |\RT|) = 2\pi z^2 \int \frac{d\kT^2}{2} \, D_1^{[0]} (z, \xi, |\kT|, |\RT|),  
\\ \label{EQ_HANG_COLL}
H_1^{\SA} (z, \xi, M_h^2) &\equiv  z^2 \int d^2\kT \, \Big[ H_1^{\SA, [0]} (z, \xi, |\kT|, |\RT|) + \frac{|\kT|}{|\RT|}  H_1^{\perp, [1]} (z, \xi, |\kT|, |\RT|) \Big] 
\\ \non
&=2\pi z^2 \int \frac{d \kT^2}{2} \, \Big[ H_1^{\SA, [0]} (z, \xi, |\kT|, |\RT|) + \frac{|\kT|}{|\RT|}  H_1^{\perp, [1]} (z, \xi, |\kT|, |\RT|) \Big] , 
}
and similarly for the barred functions.

Following Ref.~\cite{Bacchetta:2002ux}, we can expand the above DiFFs in the relative partial waves of the hadron pair system. In the center-of-mass (c.m.) frame of the pair, we can change the $\xi$ dependence to $\zeta = 2\xi -1 = a + b \cos \theta$, where $a, b$ are functions only of $M_h^2$ and $\theta$ is the angle between the direction of the back-to-back emission in the c.m. frame and the direction of $P_h$ in the target rest frame. The Jacobian of the transformation is $d\xi = |\vect{R}|/M_h d\cos \theta$. Then, we have
\al
{
\label{EQ_D_LM}
2 \frac{|\vect{R}|}{M_h} D_1 (z, \xi, M_h^2) &= D_1 (z, \cos\theta, M_h^2) = D_{1,OO} (z, M_h^2) + \cos\theta D_{1,OL}(z, M_h^2) + \ldots 
\\ \label{EQ_HANG_LM}
2 \frac{|\vect{R}|}{M_h} H_1^{\SA} (z, \xi, M_h^2) &= H_1^{\SA} (z, \cos\theta, M_h^2) = H_{1,OT}^{\SA} (z, M_h^2) + \cos\theta H_{1,LT}^{\SA} (z, M_h^2) + \ldots 
}
If we insert these expansions in \Eq{EQ_XSEC_COLL} retaining only the first nonvanishing term after integrating in $d\cos\theta$ ($d\cos\bar{\theta}$), and we further change the $y$ variable as in \Eq{EQ_Y_TH2}, then the collinear cross section can be written as
\al
{
\label{EQ_XSEC_COLL_2}
&\frac{d \sigma \Big( e^+e^- \to (h_1 h_2) (\bar{h}_1 \bar{h}_2) X \Big)}{dz d\cos\theta d\fR d M_h^2 d \bar{z} d\cos\bar{\theta} d\fRB d\bar{M}_h^2 d\cos\theta_2 } 
=  \frac{1}{4\pi^2} \frac{3 \alpha^2}{8\pi Q^2} \frac{1+\cos^2\theta_2}{4} \sum_{a,\bar{a}} e_a^2  D_{1,OO}^a (z, M_h^2) \bar{D}_{1,OO}^{\bar{a}} (\bar{z}, \bar{M}_h^2) 
\\ \non
&\hspace{1cm} \times \Bigg\{ 
1 + \cos(\fR +\fRB ) \frac{\sin^2\theta_2}{1+\cos^2\theta_2} \sin\theta \sin\bar{\theta} \frac{|\vect{R}|}{M_h} 
\frac{|\bar{\vect{R}}|}{\bar{M}_h}  
\frac{\sum_{a,\bar{a}} e_a^2 H_{1,OT}^{\SA a} (z, M_h^2) \bar{H}_{1,OT}^{\SA \bar{a}} (\bar{z}, \bar{M}_h^2)}{\sum_{a,\bar{a}} e_a^2 D_{1,OO}^a (z, M_h^2) \bar{D}_{1,OO}^{\bar{a}} (\bar{z}, \bar{M}_h^2)} 
\Bigg\} 
\\ \non
&\hspace{1cm} = \frac{1}{4\pi^2} d\sigma_0 \Big[ 1 + \cos(\fR +\fRB ) A(\cos\theta_2, \cos\theta, \cos\bar{\theta}, z, M_h^2, \bar{z}, \bar{M}_h^2) \Big] , 
}
where $\RT = \vect{R} \sin\theta$ (and similarly for $\RBT$), $d\sigma_0$ is the unpolarized cross section, and $A$ is the so-called Artru--Collins asymmetry.

The above expression is identical (up to a numerical factor) to the one used in Ref.~\cite{Courtoy:2012ry} to extract the IFF from the {\tt BELLE} experimental data for the Artru--Collins asymmetry~\cite{Vossen:2011fk}. The same IFF occurs also in the SIDIS cross section for the semi--inclusive production of hadron pairs off transversely polarized targets~\cite{Bacchetta:2002ux}, and it is used to extract the transversity distribution from a suitable single-spin asymmetry~\cite{Bacchetta:2011ip,Bacchetta:2012ty,Radici:2015mwa}. Without expanding the DiFFs in relative partial waves and by directly computing the $\cos(\fR+\fRB)$ moment of the cross section in \Eq{EQ_XSEC_COLL}, the resulting Artru--Collins asymmetry is also formally identical to that in Eq.~(23) of Ref.~\cite{Boer:2003ya} (see next section). The crucial difference is in the definition of \Eq{EQ_HANG_COLL}, namely in how the integrated IFF entering the asymmetry is built in terms of unintegrated DiFFs. Starting from the correct cross section of \Eq{EQ_XSEC_V1}, the expression in  \Eq{EQ_HANG_COLL} (multiplied by $|\RT|$) is now consistent with the definition of IFF entering the azimuthal asymmetry in the SIDIS cross section~\cite{Matevosyan:2017uls} (see also Ref.~\cite{Bacchetta:2003vn}). The same consistency could not be achieved from the cross section in Eq.~(19) of Ref.~\cite{Boer:2003ya}. Thus, the discrepancy is indeed resolved.

\section{The azimuthal asymmetries}
\label{SEC_EE_ASYMM}

In this section, we will review and discuss the azimuthal asymmetries that allow to extract the IFF and the helicity dependent DiFF from the cross section listed in \Eq{EQ_XSEC_V2}. For this purpose, we define the average of an arbitrary function $\mathcal{I}$ as
\al
{
\label{EQ_XSEC_AVERAGE}
\langle \mathcal{I} \rangle \equiv
  \int d\xi & \int d\bar{\xi} \int d\fR \int d\fRB \int d^2 \qT  \ \mathcal{I} 
\  \frac{d \sigma \Big( e^+e^- \to (h_1 h_2) (\bar{h}_1 \bar{h}_2) X \Big)}{d^2 \qT dz d\xi d\fR d M_h^2 d \bar{z} d\bar{\xi} d\fRB d \bar{M}_h^2 dy }.
}
We first calculate the integral of the unweighted cross section, that appears as denominator in all of the azimuthal asymmetries. Following the same steps leading to \Eq{EQ_XSEC_COLL}, we have
\al
{
\label{EQ_XSEC_INT_UPOL}
 \langle 1 \rangle &= \int d\xi \int d\bar{\xi} \int d\fR \int d\fRB \int d^2 \qT  \  
    \frac{d \sigma \Big( e^+e^- \to (h_1 h_2) (\bar{h}_1 \bar{h}_2) X \Big)}{d^2 \qT dz d\xi d\fR d M_h^2 d \bar{z} d\bar{\xi} d\fRB  d\bar{M}_h^2 dy }
 \\ \non
& = \frac{3 \alpha^2}{\pi Q^2}  A(y) \sum_{a,\bar{a}} e_a^2 
 D_1^a(z, M_h^2) \bar{D}_1^{\bar{a}}(\bar{z}, \bar{M}_h^2),
}
where
\al
{
 D_1^a(z, M_h^2) = \int d\xi \, D_1^a (z, \xi, M_h^2) ,
}
and $D_1 (z, \xi, M_h^2)$ is given in \Eq{EQ_D_COLL} (and similarly for $\bar{D}_1^{\bar{a}}$).

\subsection{Artru-Collins asymmetry}
\label{SUBSEC_ARTU_COLLINS}

In Ref.~\cite{Boer:2003ya}, the Artru-Collins asymmetry is defined as
\al
{
\label{EQ_ASYMM_INT_IFF}
 A(y, z , \bar{z}, M_h^2, \bar{M}_h^2) = \frac{\langle \cos(\fR + \fRB ) \rangle}{\langle 1 \rangle} .
}
Following the same steps leading to \Eq{EQ_XSEC_COLL}, we have
\al
{
\label{EQ_IFF_MOM}
\langle \cos(\fR + \fRB) \rangle
= \frac{3 \alpha^2}{2\pi Q^2}  B(y) \sum_{a,\bar{a}} e_a^2  
H_1^{\SA a} (z, M_h^2) \bar{H}_1^{\SA \bar{a}} (\bar{z}, \bar{M}_h^2),
}
where
\al
{
\label{EQ_HANG_COLL_2}
H_1^{\SA} (z, M_h^2) = \int d\xi \, \frac{|\RT|}{M_h} H_1^{\SA} (z, \xi, M_h^2), 
}
with $H_1^{\SA} (z, \xi, M_h^2)$ given in \Eq{EQ_HANG_COLL}  (and similarly for $\bar{H}_1^{\SA \bar{a}}$).

Finally, the Artru--Collins asymmetry results
\al
{
\label{EQ_ART_COL_ASYMM}
A(y, z , \bar{z}, M_h^2, \bar{M}_h^2) = \frac{1}{2}
\Bigg[
 B(y)  \sum_{a,\bar{a}} e_a^2  
  H_1^{\SA a}(z, M_h^2) \bar{H}_1^{\SA \bar{a}}(\bar{z},\bar{M}_h^2)
\Bigg]
\Bigg[
 A(y)  \sum_{a,\bar{a}} e_a^2 
  D_1^a(z, M_h^2) \bar{D}_1^{\bar{a}}(\bar{z}, \bar{M}_h^2)  
\Bigg]^{-1} , 
}
which is identical to Eq.~(23) of Ref.~\cite{Boer:2003ya}, but now $H_1^{\SA a} (z, M_h^2)$ is given by \Eq{EQ_HANG_COLL_2} consistently with the definition entering the azimuthal asymmetry in the SIDIS cross section~\cite{Matevosyan:2017uls} (and similarly for $\bar{H}_1^{\SA \bar{a}}(\bar{z},\bar{M}_h^2)$).

\subsection{The asymmetry for the helicity dependent DiFF}
\label{SUBSEC_GPERP}

Another important consequence of the new expression for the cross section in \Eq{EQ_XSEC_V2} is that the so-called longitudinal jet handedness azimuthal asymmetry, suggested in Ref.~\cite{Boer:2003ya} to address the helicity dependent DiFF, identically vanishes. This asymmetry is defined as
\al
{
\label{EQ_GPERP_ASYCOS2}
 A^{\Rightarrow}(y, z , \bar{z}, M_h^2, \bar{M}_h^2) = \frac{\langle \cos( 2(\fR - \fRB) ) \rangle}{\langle 1 \rangle}.
}

The contributions to $\langle \cos( 2(\fR - \fRB) \rangle$ from terms in \Eq{EQ_XSEC_V2} involving $B(y)$ vanish, which is easy to check using similar steps to those used in the derivations of \Eq{EQ_XSEC_COLL}, where we quickly end up with an expression  multiplied by
\al
{
\int d \fR \int d \fRB \cos(2(\fR-\fRB)) \cos(\fR+\fRB) = 0.
}

The only remaining contribution is by the last term in \Eq{EQ_XSEC_V2}, which we can again transform to a much simpler form by redefining $\fk\to \fKR$, $\fkB\to \fKRB$ after integrating upon $d\qT$:
\al
{
\label{EQ_GPERP_MOM_COS2}
\langle &\cos(2(\fR - \fRB)) \rangle
 =  -  \frac{3 \alpha^2}{ \pi Q^2}   A(y)  
\sum_{a,\bar{a}} e_a^2  
\int d\xi  \int d\bar{\xi} \int d\fR \int d\fRB 
\ \cos(2(\fR - \fRB))
\\ \non
&\times z^2 \bar{z}^2 \int d\kT \int d\kBT
 \sin(\fk) \sin(\fkB )
\frac{|\kT| |\RT|}{M_h^2}   G_1^{\perp a}\Big(z,\xi, |\kT|, |\RT|, \cos(\fK)\Big)
\frac{|\kBT| |\RBT|}{\bar{M}_h^2} \bar{G}_1^{\perp \bar{a}}\Big(\bar{z},\bar{\xi}, |\kBT|, |\RBT|, \cos(\fkB)\Big)
\\ \non
& = 0  .
}
Thus, the asymmetry of \Eq{EQ_GPERP_ASYCOS2} identically vanishes.
In fact, any moment of the cross section that depends only on angles $\fR$ and $\fRB$ would get no contribution from the terms involving $G_1^\perp$, as can readily be seen from the derivation in \Eq{EQ_GPERP_MOM_COS2} since the integration upon $d\qT$ already yields a zero. 

It is interesting to investigate if there is a specific moment that allows to single out the helicity dependent DiFF $G_1^\perp$. If we include in the weight information on $|\qT|$, following the same steps as before for example we get
\al
{
\label{EQ_GPERP_MOM_QTCOS}
\langle \qT^2 \cos(\fR - \fRB) \rangle
 =  & \frac{3 \alpha^2}{ \pi Q^2}  A(y)  
\sum_{a,\bar{a}} e_a^2  M_h \bar{M}_h 
\Bigg\{  2 D_1^{a, [1], (1/2)} (z, M_h^2) \bar{D}_1^{\bar{a}, [1], (1/2)} (\bar{z}, \bar{M}_h^2) 
\\ \non
&\hspace{1cm}
 - \ \Big( G_1^{\perp a, [0], (1)} (z, M_h^2) - G_1^{\perp a, [2], (1)} (z, M_h^2) \Big) \  
\Big(  \bar{G}_1^{\perp \bar{a}, [0], (1)} (\bar{z}, \bar{M}_h^2) - \bar{G}_1^{\perp \bar{a}, [2], (1)} (\bar{z}, \bar{M}_h^2) \Big) \Bigg\}
\\ \non
& \hspace{-0.3cm} \equiv  
\frac{3 \alpha^2}{\pi Q^2}  A(y) \sum_{a,\bar{a}} e_a^2  M_h \bar{M}_h 
\Bigg\{  2 D_1^{a, [1], (1/2)} (z, M_h^2) \bar{D}_1^{\bar{a}, [1], (1/2)} (\bar{z}, \bar{M}_h^2) 
- G_1^{\perp a} (z, M_h^2)
 \bar{G}_1^{\perp \bar{a}} (\bar{z}, \bar{M}_h^2)  \Bigg\} , 
}
where
\al
{
\label{EQ_KTMOMD}
D_1^{[n], (p)} (z, M_h^2) &\equiv z^2 \int d^2\kT \, \left( \frac{\kT^2}{2 M_h^2}\right)^p  \int d\xi \, D_1^{[n]} (z, \xi, |\kT|, |\RT|),
\\
\label{EQ_KTMOMGPERP}
G_1^{\perp, [n], (p)} (z, M_h^2) &\equiv z^2 \int d^2\kT \, \left( \frac{\kT^2}{2 M_h^2}\right)^p  \int d\xi \, \frac{|\RT|}{M_h} \, G_1^{\perp, [n]} (z, \xi, |\kT|, |\RT|), 
\\
\label{EQ_INT_GPERP}
G_1^{\perp} (z, M_h^2) &\equiv G_1^{\perp, [0], (1)} (z, M_h^2) - G_1^{\perp, [2], (1)} (z, M_h^2) , 
}
are $\kT^2-$moments of order $p$ of the Fourier cosine moments of order $n$ of the involved DiFFs (and similarly for the barred functions). Note, that this definition of $G_1^{\perp} (z, M_h^2)$ is different than that in Ref~\cite{Boer:2003ya}. Therefore, weighing the cross section with a function of $\fR, \fRB$ and $\qT^2$ is not enough to isolate its contribution coming from the helicity dependent DiFF. 

Such new weight has been recently proposed in Ref.~\cite{Matevosyan:2017liq}, that also involves the azimuthal angle $\fq = \vf_1 + \pi$ of $\qT$ to exactly cancel out the contributions from the unpolarized term in the cross section:
\al
{
&\left \langle \qT^2 \Big( 3 \sin (\fq - \fR)  \sin(\fq - \fRB) + \cos (\fq - \varphi_R)  \cos (\fq - \varphi_{\bar{R}}) \Big) \right \rangle  
\\ \non
&\hspace{0.7cm} = \left \langle \qT^2 \Big( 2 \cos (\fR - \fRB) - \cos(2 \vf_1 - \fR - \fRB)  \Big) \right \rangle 
\\ \non
&\hspace{0.7cm}
= \frac{12 \alpha^2}{\pi  Q^2}  A(y)  \sum_{a,\bar{a}} e_a^2  M_h \bar{M}_h 
\ G_1^{\perp a} (z, M_h^2)  \  
\ \bar{G}_1^{\perp \bar{a}} (\bar{z}, \bar{M}_h^2) , 
}
where $G_1^{\perp } (z, M_h^2)$ is defined in \Eq{EQ_INT_GPERP} (and similarly for $\bar{G}_1^{\perp} (\bar{z}, \bar{M}_h^2)$ ).

Finally, it is worth noticing that since $\langle \qT^2 \cos (\fR - \fRB) \rangle \neq 0$ and $\langle \cos (\fR - \fRB) \rangle = 0$, the latter moment can contain terms that survive the integration upon $\fq$ but vanish because of the integration upon the modulus $|\qT|$. If we perform all the integrations indicated in \Eq{EQ_XSEC_AVERAGE} except for the one upon $d|\qT|$, the only surviving contribution is (see Appendix~\ref{SEC_APPEND_A})
\al
{
\label{EQ_GPERP_MOM_COS}
\langle \cos(\fR - \fRB) \rangle (\qT^2)
&= \frac{3 \alpha^2}{ \pi Q^2}  A(y)  \sum_{a,\bar{a}} e_a^2  \int d\fq \, {\cal F}_1^a (\qT^2, z, \bar{z}, \RT^2, \RBT^2) 
\\ \non
&= \frac{3 \alpha^2}{\pi Q^2}  A(y)  \sum_{a,\bar{a}} e_a^2 \, 2 \pi \, {\cal F}_1^a \neq 0
}
where
\al
{
\label{EQ_GP_MOM}
{\cal F}_1^a (\qT^2, z, \bar{z}, \RT^2, \RBT^2) &= \int d^2\kT \int d^2 \kBT \, \delta (\kT + \kBT - \qT) \cos (\fk - \fkB) 
\\ \non
&\times \Bigg\{ \int d\xi D_1^{a, [1]} (z, \xi, |\kT|, |\RT|) \int d\bar{\xi} \bar{D}_1^{\bar{a}, [1]} (\bar{z}, \bar{\xi}, |\kBT|, |\RBT|) 
\\ \non
&\hspace{0.7cm} - \frac{1}{4}  \int d\xi \frac{|\kT| |\RT|}{M_h^2} \Big( G_1^{\perp  a, [0]} (z, \xi, |\kT|, |\RT|) - G_1^{\perp  a, [2]} (z, \xi, |\kT|, |\RT|) \Big)  
\\ \non
&\hspace{2cm} \times 
 \int d\bar{\xi} \frac{|\kBT| |\RBT|}{\bar{M}_h^2} \Big( \bar{G}_1^{\perp  \bar{a}, [0]} (\bar{z}, \bar{\xi}, |\kBT|, |\RBT|) - \bar{G}_1^{\perp  \bar{a}, [2]} (\bar{z}, \bar{\xi}, |\kBT|, |\RBT|) \Big)  \Bigg\} .
}

If $\langle \cos (\fR - \fRB) \rangle = 0$ vanishes because of the integration upon the modulus $|\qT|$, it means that this moment, when considered as a function of $\qT^2$, must have a node. Indeed, some preliminary measurements from the {\tt BELLE} collaboration indicate a non vanishing $\langle \cos (\fR - \fRB) \rangle$ which could be due to the limited coverage  in $\qT^2$~\cite{Anselm:2013}. However, it is not evident which combination of moments of DiFFs in \Eq{EQ_GP_MOM} is responsible for a node in \Eq{EQ_GPERP_MOM_COS}. In principle, both terms could contribute in changing the sign of $\langle \cos (\fR - \fRB) \rangle$ because the Fourier cosine moment $D_1^{[1]}$ is not necessarily a positive definite function. 
\end{widetext}

\section{Conclusions}
\label{SEC_CONCLUSIONS}

 The DiFFs provide a very rich source of information concerning the hadronization process. Moreover, in recent years they have been used to explore the structure of the nucleon using two-hadron semi-inclusive electroproduction. The information about the DiFFs extracted from the two back-to-back hadron pair semi-inclusive production in $e^+e^-$ annihilation plays an absolutely vital role in these studies. The fully unintegrated cross section for this process and the relevant azimuthal asymmetries for accessing the different DiFFs were first derived in Ref.~\cite{Boer:2003ya}. 
 
We recently observed in Ref.~\cite{Matevosyan:2017uls} that the integrated IFF built from the DiFFs entering the unintegrated SIDIS cross section is apparently different from the one that is built from the corresponding unintegrated cross section for $e^+e^-$ annihilation obtained in Ref.~\cite{Boer:2003ya}. In this work we rederived these quantities following the same kinematic setup of Ref.~\cite{Boer:2003ya}. In Sec.~\ref{SUBSEC_EE_XSEC}, we found a mistake in the definition of the kinematics that impacts a subset of terms in the cross section having significant implications for the relevant asymmetries. The most important result derived in Sec.~\ref{SUBSEC_ARTU_COLLINS} is that with the corrected cross section the apparent discrepancy between the definitions of the integrated IFF in terms of unintegrated DiFFs occurring in the SIDIS and $e^+ e^-$ cross sections is resolved. Although the procedure used in the extraction of the transversity PDF using the dihadron method in Refs.~\cite{Bacchetta:2011ip,Bacchetta:2012ty,Radici:2015mwa} is formally correct, it is nevertheless important to have a consistent underlying formalism, which has been established here. 
 
The second important result, derived in Sec.~\ref{SUBSEC_GPERP}, is that that azimuthal asymmetry previously proposed for accessing the helicity dependent DiFF $G_1^\perp$ actually vanishes. The reason is the complete decoupling of the quark and antiquark transverse momenta in these asymmetries, as a consequence of which the modulations of their respective hadron productions are lost.  This naturally explains the absence of the corresponding signal in the recent analysis at {\tt BELLE}~\cite{Abdesselam:2015nxn, Vossen:2015znm}. Further, we discussed the azimuthal asymmetry recently proposed in Ref.~\cite{Matevosyan:2017liq} that allows to access $G_1^\perp$. We have also analyzed another azimuthal asymmetry based on the relative azimuthal orientation of the planes containing the two back-to-back hadron pair momenta. Interestingly, this asymmetry vanishes independently of the various angular integrations, because it displays a node as a function of the size of the imbalance between the transverse momenta of the two back-to-back jets. As a consequence, incomplete integration on the imbalance size would generate a nonvanishing result, as well as including also the imbalance size as an additional weight.

  An important next step is to extend these calculations to beyond the leading-twist contributions, both in the kinematic factors and the DiFFs themselves. The need for this is motivated by the upcoming and planned next generation experiments.  

\vspace{-0.6cm}

\section*{ACKNOWLEDGEMENTS}

The work of H.H.M. and A.W.T. was supported by the Australian Research Council through the ARC Centre of Excellence for Particle Physics at the Terascale (CE110001104), and by the ARC  Discovery Project No. DP151103101, as well as by the University of Adelaide. A.K. was supported by A.I. Alikhanyan National Science Laboratory (YerPhI) Foundation, Yerevan, Armenia. The work of A.B. and M.R. is supported by the European Research Council (ERC) under the European Union's Horizon 2020 research and innovation program (grant agreement No. 647981, 3DSPIN). A. Courtoy wishes to thank the CICOPS of the University of Pavia for the support received during the preparation of this article.

\begin{widetext}

\appendix
\vspace{-2.5cm}
\section{COSINE MOMENT OF RELATIVE ORIENTATION OF HADRON PAIRS PLANES}
\label{SEC_APPEND_A}

By performing all the integrations indicated in \Eq{EQ_XSEC_AVERAGE} except for the one upon $d\qT$, the $\langle \cos (\fR - \fRB) \rangle$ moment becomes
\al
{
\label{EQ_FULL_MOMENT}
\langle \cos(\fR - &\fRB) \rangle (\qT) =  \frac{3 \alpha^2}{\pi Q^2}  \sum_{a,\bar{a}} e_a^2   \int d\xi \int d\bar{\xi} 
 \int d^2\kT \int d^2 \kBT \delta (\kT + \kBT - \qT) 
\\ \non
&\times \int d\fR d\fRB \cos(\fR - \fRB) \sum_{n,m}  \frac{\cos n(\fK-\fR)}{\pi (1+\delta_{0,n})}  \frac{\cos m(\fkB-\fRB)}{\pi (1+\delta_{0,m})} 
\\ \non
&\times
 \Bigg\{
A(y) D_1^{a, [n]}  \bar{D}_1^{\bar{a}, [m]} 
\\ \non 
&\hspace{0.5cm} + B(y) \cos(\fK +\fkB)  \frac{|\kT|}{M_h}  H_1^{\perp a, [n]} \frac{|\kBT|}{\bar{M}_h} \bar{H}_1^{\perp \bar{a}, [m]} +
B(y) \cos(\fR +\fRB ) \frac{|\RT|}{M_h} H_1^{\SA a, [n]} \frac{|\RBT|}{\bar{M}_h} \bar{H}_1^{\SA \bar{a}, [m]}
\\ \non
&\hspace{0.5cm} +  B(y) \cos(\fk + \fRB ) \frac{|\kT|}{M_h} H_1^{\perp a, [n]} \frac{|\RBT|}{\bar{M}_h} \bar{H}_1^{\SA \bar{a}, [m]} + 
B(y) \cos(\fkB  + \fR) \frac{|\RT|}{M_h}  H_1^{\SA a, [n]}  \frac{|\kBT|}{\bar{M}_h} \bar{H}_1^{\perp \bar{a}, [m]} 
\\ \non
&\hspace{0.5cm} -  A(y) \sin(\fk - \fR ) \sin(\fkB - \fRB ) \frac{|\kT|  |\RT|}{M_h^2} G_1^{\perp a, [n]}  \frac{|\kBT| |\RBT|}{\bar{M}_h^2} \bar{G}_1^{\perp \bar{a}, [m]}
\Bigg\}  
}
\al{
\non
&= \frac{3 \alpha^2}{\pi Q^2}  \sum_{a,\bar{a}} e_a^2  \int d^2\kT \int d^2 \kBT \delta (\kT + \kBT - \qT) 
\Bigg\{ A(y) \cos(\fk - \fkB ) \int d\xi D_1^{a, [1]} \int d\bar{\xi} \, \bar{D}_1^{\bar{a}, [1]}  
\\ \non
&\quad + B(y)  \cos(\fk - \fkB )  \cos(\fk + \fkB )  \int d\xi \frac{|\kT|}{M_h} H_1^{\perp a, [1]} \int d\bar{\xi} \, \frac{|\kBT|}{\bar{M}_h}  \bar{H}_1^{\perp \bar{a}, [1]} 
\\ \non
&\quad + B(y) \frac{1}{2} \Bigg[ \cos 2\fk \int d\xi \frac{|\RT|}{M_h} H_1^{\SA a,  [2]} \int d\bar{\xi} \frac{|\RBT|}{\bar{M}_h} \bar{H}_1^{\SA \bar{a}, [0]} + 
 \cos 2\fkB \int d\xi \frac{|\RT|}{M_h} H_1^{\SA a, [0]} \int d\bar{\xi} \frac{|\RBT|}{\bar{M}_h} \bar{H}_1^{\SA \bar{a}, [2]}
\\ \non
&\hspace{0cm} + \int d\xi \frac{|\kT|}{M_h} H_1^{\perp a, [1]} \int d\bar{\xi} \frac{|\RBT|}{\bar{M}_h} \Big( \cos 2\fk \bar{H}_1^{\SA \bar{a}, [0]}  + \cos 2\fkB \bar{H}_1^{\SA \bar{a}, [2]} \Big) 
\\ \non
&\hspace{2cm}
+ \int d\bar{\xi} \frac{|\kBT|}{\bar{M}_h} \bar{H}_1^{\perp \bar{a}, [1]} \int d\xi \frac{|\RT|}{M_h} \Big(  \cos 2\fkB H_1^{\SA a, [0]}  + \cos 2\fk H_1^{\SA a, [2]} \Big) \Bigg] 
\\ \non
&\quad - A(y) \frac{1}{4} \cos(\fk - \fkB ) \int d\xi \frac{|\kT| |\RT|}{M_h^2} \Big( G_1^{\perp a, [0]} - G_1^{\perp a, [2]} \Big) \int d\bar{\xi} \frac{|\kBT| |\RBT|}{\bar{M}_h^2} \Big( \bar{G}_1^{\perp \bar{a}, [0]} - \bar{G}_1^{\perp \bar{a}, [2]} \Big) 
\Bigg\} 
\\ \non
&\equiv \frac{3 \alpha^2}{\pi Q^2}   \sum_{a,\bar{a}} e_a^2  \int d^2\kT \int d^2 \kBT \, \delta (\kT + \kBT - \qT) 
\\ \non
&\times \Bigg\{ A(y) \cos (\fk - \fkB) F_1^a + B(y) \cos (\fk - \fkB) \cos (\fk + \fkB) F_2^a  + \frac{B(y)}{2} \cos 2\fk F_3^a + \frac{B(y)}{2} \cos 2\fkB F_4^a \Bigg\} , 
}
where
\al
{
\label{F1def}
F_1^a (z, \bar{z}, \kT^2, \kBT^2, \RT^2, \RBT^2) &=   
\int d\xi  D_1^{a, [1]} \int d\bar{\xi}  \bar{D}_1^{\bar{a}, [1]} 
\\ \non
&\quad - \frac{1}{4}  \int d\xi \frac{|\kT| |\RT|}{M_h^2} \Big( G_1^{\perp  a, [0]} - G_1^{\perp  a, [2]} \Big)  
                      \int d\bar{\xi} \frac{|\kBT| |\RBT|}{\bar{M}_h^2} \Big( \bar{G}_1^{\perp  \bar{a}, [0]} - \bar{G}_1^{\perp  \bar{a}, [2]} \Big) 
\\ \non
&\equiv |\kT| |\kBT| \Big[ 
F_1^D (z, \kT^2, \RT^2) \bar{F}_1^{\bar{D}} (\bar{z}, \kBT^2, \RBT^2) + F_1^G (z, \kT^2, \RT^2) \bar{F}_1^{\bar{G}} (\bar{z}, \kBT^2, \RBT^2) \Big]
\\ \label{F2def}
F_2^a (z, \bar{z}, \kT^2, \kBT^2, \RT^2, \RBT^2) &= 
\int d\xi \frac{|\kT|}{M_h} H_1^{\perp a, [1]} \int d\bar{\xi} \frac{|\kBT|}{\bar{M}_h}  \bar{H}_1^{\perp \bar{a}, [1]} 
\equiv |\kT| |\kBT|  F_2^H (z, \kT^2, \RT^2) \bar{F}_2^{\bar{H}} (\bar{z}, \kBT^2, \RBT^2)
\\ \label{F3def}
F_3^a (z, \bar{z}, \kT^2, \kBT^2, \RT^2, \RBT^2) &= 
\int d\xi \frac{|\RT|}{M_h} H_1^{\SA a, [2]} \int d\bar{\xi} \Big( \frac{|\RBT|}{\bar{M}_h} \bar{H}_1^{\SA \bar{a}, [0]} + \frac{|\kBT|}{\bar{M}_h} \bar{H}_1^{\perp \bar{a}, [1]} \Big) + \int d\xi \frac{|\kT|}{M_h} H_1^{\perp a, [1]} \int d\bar{\xi} \frac{|\RBT|}{\bar{M}_h} \bar{H}_1^{\SA \bar{a}, [0]} 
\\ \label{F4def}
F_4^a (z, \bar{z}, \kT^2, \kBT^2, \RT^2, \RBT^2) &=  
\int d\bar{\xi} \frac{|\RBT|}{\bar{M}_h} \bar{H}_1^{\SA \bar{a}, [2]} \int d\xi \Big( \frac{|\RT|}{M_h} H_1^{\SA a, [0]} + \frac{|\kT|}{M_h} H_1^{\perp a, [1]} \Big) + \int d\bar{\xi} \frac{|\kBT|}{\bar{M}_h} \bar{H}_1^{\perp \bar{a}, [1]} \int d\xi \frac{|\RT|}{M_h} H_1^{\SA a, [0]}  .
}

The integral on $d\fq$ of the $\langle \cos (\fR - \fRB) \rangle (\qT)$ moment in \Eq{EQ_FULL_MOMENT} is nonzero. In fact, the first term of the last line gives
\al
{
&\int d\fq \int d^2\kT \int d^2 \kBT \, \delta (\kT + \kBT - \qT)  \cos (\fk - \fkB ) \, F_1^a (z, \bar{z}, \kT^2, \kBT^2, \RT^2, \RBT^2) 
\\ \non 
&= \int d\fq \int d^2\kT \int d^2 \kBT \frac{1}{(2\pi)^2} \int d^2\b \, e^{i \b \cdot (\qT - \kT - \kBT)} \kT \cdot \kBT \Big( F_1^D \bar{F}_1^{\bar{D}}  + F_1^G \bar{F}_1^{\bar{G}} \Big)  
\\ \non
&= \int d\fq \frac{1}{(2\pi)^2} \int d^2\b \, e^{i \b \cdot \qT} (- 4 \b^2) \Bigg[ \frac{\partial}{\partial \b^2} \hat{F}_1^D (z, \b^2, \RT^2) \frac{\partial}{\partial \b^2} \hat{\bar{F}}_1^{\bar{D}} (\bar{z}, \b^2, \RBT^2)  + \frac{\partial}{\partial \b^2} \hat{F}_1^G (z, \b^2, \RT^2) \frac{\partial}{\partial \b^2} \hat{\bar{F}}_1^{\bar{G}} (\bar{z}, \b^2, \RBT^2)  \Bigg]
\\ \non
&=  \int d\fq \, {\cal F}_1^a (\qT^2, z, \bar{z}, \RT^2, \RBT^2) = 2\pi {\cal F}_1^a \neq 0 ,
}
where $\hat{F}_1^D, \hat{F}_1^G$ ($\hat{\bar{F}}_1^{\bar{D}}, \hat{\bar{F}}_1^{\bar{G}}$) are the inverse Fourier transforms of $F_1^D, F_1^G$ ($\bar{F}_1^{\bar{D}}, \bar{F}_1^{\bar{G}}$), respectively. 

Following similar steps, it is easy to verify that 
\al
{
&\int d\fq \int d^2\kT \int d^2 \kBT \, \delta (\kT + \kBT - \qT)  \cos (\fk - \fkB) \cos (\fk + \fkB) \, F_2^a (z, \bar{z}, \kT^2, \kBT^2, \RT^2, \RBT^2) 
\\ \non
&= \int d\fq \int d^2\kT \int d^2 \kBT \frac{1}{(2\pi)^2} \int d^2\b \, e^{i \b \cdot (\qT - \kT - \kBT)} \kT \cdot \kBT \cos (\fk + \fkB) \, F_2^H \bar{F}_2^{\bar{H}}
\\ \non
&= 32 \int d\fq \Big\{ (q_y^2 - q_x^2) \frac{\partial^2}{\partial (\qT^2)^2} {\cal F}_2^a (\qT^2, z, \bar{z}, \RT^2, \RBT^2) 
\\ \non
&\hspace{2cm} + 8 \Big[ 3 (q_x^2 - q_y^2) \frac{\partial^3}{\partial (\qT^2)^3} + (q_x^4 - q_y^4) \frac{\partial^4}{\partial (\qT^2)^4} \Big] {\cal F}_2^{\prime  a} (\qT^2, z, \bar{z}, \RT^2, \RBT^2)  \Big\} = 0 ,
\\[1cm] 
&\int d\fq \int d^2\kT \int d^2 \kBT \, \delta (\kT + \kBT - \qT) \cos 2\fk \, F_3^a (z, \bar{z}, \kT^2, \kBT^2, \RT^2, \RBT^2)
\\ \non
&=  16 \int d\fq (q_x^2 - q_y^2)  \frac{\partial^2}{\partial (\qT^2)^2}  \, {\cal F}_3^a (\qT^2, z, \bar{z}, \RT^2, \RBT^2) = 0 ,
}
and similarly for $F_4^a (z, \bar{z}, \kT^2, \kBT^2, \RT^2, \RBT^2)$. 
\\
\end{widetext}

\bibliography{fragment}

\end{document}